# EFFECTIVENESS OF RECONFIGURABLE INTELLIGENT SURFACE IN MULTIPATH FADING CHANNEL

*Hasnul Hashim*

*Electrical and Electronic Engineering Programme Area, Tungku Link, Universiti Teknologi Brunei, Brunei Darussalam*
*\*hasnul.hashim@utb.edu.bn*

## Abstract

A method of simulating a single-input single-output reconfigurable intelligent surface (RIS) assisted channel is presented using three channel black boxes to represent the direct signal path, the transmit path to the RIS and the reflected path from the RIS. The complex coefficients for each channel box is obtained by ray tracing in a scenario with geographic terrain information that also contains approximate building shapes. The electrical characteristics of the ground and building walls were also accounted for in the ray tracing function. Simulations were conducted with reflected rays only and reflected rays together with diffracted rays. The received power exhibits variations typical of multipath fading environments. In the best locations, the RIS-assisted channel simulation result agrees well with theoretical models, the performance increasing by the RIS size squared as the number of RIS elements is increased. In the simplified theoretical model where the transmitter and receiver are inline and the RIS orthogonal but much closer than the distance between the former elements, the simulation results also corroborate best deployment close the transmitter or the receiver with a U-shaped drop between them.

## 1   Introduction

Reconfigurable intelligent surface is one of the technologies being considered for 6G wireless communications. In the literature, it is also referred to as intelligent reflecting surface [1], [2], the, overarching, intelligent surface [3], software-controlled metasurface [4] and reconfigurable metasurface [5]. One of the purported benefits of RIS is the ability to control [6], [7], [8] some aspects of radio wave propagation in typical wireless environments that are usually uncontrolled. However, the RIS benefits are tempered by challenging questions [9].

Such wireless communications are often hampered by rapid variations of the signal, i.e. multipath fading. In [10], the size of the RIS needed to compete with a relaying scheme is considered in a frequency flat fading channel. In [11], the multipath fading for mobile users is claimed to be eliminated by tuning the RIS in real time. A similar claim was made in [12] with experimental measurement at 35 GHz. In [13] and [14], the RIS is a spatial equalizer to mitigate ISI in multipath communication.

The multipath works referenced above employed models based on traditional linear systems. Many small-scale multipath fading models for RIS have been devised. A brief survey can be found in [15]. A widely used RIS model is given in [16] based on the transmission modes of impinging electromagnetic waves (EM). Another physics-based model was developed in [17] based on wave optics and multipath statistics. In [18], a geometry-based stochastic channel model was used for a RIS-assisted channel.

In contrast to these RIS models, our work uses the method of ray tracing [19] to model multipath fading with RIS. The method has been used for antenna arrays [20] which are related to RIS. The geometry of the environment is taken into account as in [18] but the ray tracing element is deterministic and also takes into account electrical characteristics of the channel. The ray tracing method is a physics-based approach in the category of computational EM. Unlike [16] and [17], the RIS itself is simply modelled by sets of complex numbers that represent the steering and beamforming ability of the device [8].

In the following sections, the RIS-assisted channel model and a ray tracing simulation method for the channel is described before the simulation results are presented in the section following it. For the remainder of this paper, the terms transmitter, base station (BS) and gNodeB (gNB) are used interchangeably and refer to the same device.

## 2.   RIS-Assisted Channels

*2.1 Single-Input Single-Output RIS Channel*

A typical single-input single-output RIS-assisted channel used in this work is shown in Fig. 1 where the base station is in line-of-sight of the user equipment (UE). The channel response of the line-of-sight link is $h_s$. Scenarios where the UE is not in line-of-sight of the BS, primarily due to building blockage are also considered. The RIS consists of *N* metaatoms, each capable of reflecting an incident signal and controlling the phase shift of the reflected signal. Typically, the metaatoms are arranged in a rectangular array with a spacing of λ/2 where λ is the carrier wavelength.

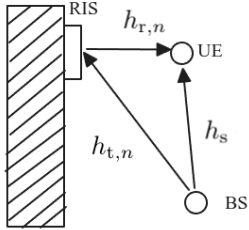

Fig. 1 Base station in line-of-sight of the UE with a RIS mounted on the side of a building as viewed from a high elevation.

The total channel response *h* of a RIS-assisted channel [8] is given by

$$h = h_s + \sum_{n=1}^{N} h_{r,n} e^{j\psi_n} h_{t,n} \quad (1)$$

where $h_{t,n}$ is the channel response from the BS to the *n*-th metaatom, $h_{r,n}$ is the channel response from the *n*-th metaatom to the UE and $\psi_n$ is the phase shift of the *n*-th metaatom. The set of $\psi_n$ are called the RIS coefficients. It can be shown [8] that the *optimal* RIS coefficients that maximises the capacity of the channel are obtained when the phase shifts satisfy the equation

$$\psi_n = \sphericalangle h_{r,n} - \sphericalangle h_{t,n} + 2\pi k_n \quad (2)$$

for some integer $k_n$ that wraps the the *n-th* phase shift between -π and π.

## 2.2 RIS Simulation Methodology

In order to study the RIS within a multipath fading environment, the simulation scheme shown in Fig. 2 is proposed where each component of the channel in Fig. 1 is modelled using the method of ray tracing.

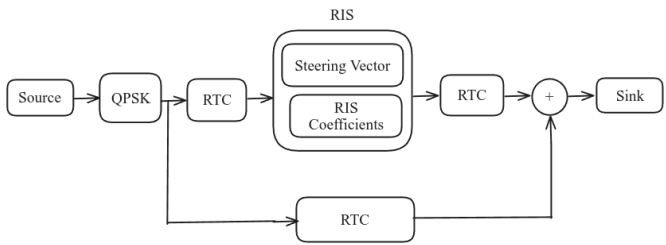

Fig. 2 Block diagram of the simulation scheme using ray traced channels (RTC) to model the RIS.

The direct channel $h_s$ is simulated first, using the bottom signal path, followed by the cascade of the $h_{r,n}$, RIS and $h_{t,n}$ channel in the top path. The results are summed under the linearity assumption given by (1). The RIS coefficients are computed using (2) with the angles of the incoming and outgoing rays to and from the RIS obtained from the ray-traced channels. The additional phase delays of the RIS array are accounted for using a steering vector [21].

Unless stated otherwise, the general simulation parameters are shown Table 1. In addition, the antennas used by the UE and base station are isotropic. The RIS antenna elements were configured as half-isotropic. Rays potentially arriving at the RIS from the anti-normal side are discarded. This is equivalent to the assumption of placing the RIS on a wall with the normal side pointing towards the desired propagation environment and the anti-normal side completely inaccessible to the rays.

For convenience, the propagation environment geometry and material composition were extracted from a geographical database. Specifically, it was obtained from OpenStreetMap and the region of interest is derived from a square area in a suburban city on flat terrain. The coordinates of the southwest corner of this region is given in Table 1.

The simulations were solely conducted on the Matlab platform using version R2025a. In particular, the 'comm.RayTracingChannel' object and 'raytrace()' functions were heavily used. The ray tracing algorithm used is the shooting and bouncing rays method. Some of the parameters used for the 'raytrace()' function are given in Table 1. Optional parameters were left at their default values. In particular, polarization is not modelled explicitly but left at the default values used by Matlab for all ray reflection and diffraction computations. (Matlab asssumes linear polarization and equally weights the horizontal and vertical components of the Jones vector.)

Table 1 General Simulation Parameters

| Parameter | Value |
| --- | --- |
| Carrier frequency | 28 GHz |
| Transmitter power | 30 dBm |
| Transmitter height | 5 m |
| UE antenna height | 1 m |
| RIS height | 5 m |
| RIS array elevation | -3° |
| Building material | Brick [22] |
| Terrain material | Concrete [22] |
| Southwest corner latitude | 3.07351° |
| Southwest corner longitude | 101.58633° |
| RIS array arrangement | square |
| *N* | 1024 |

## 2.3 Simulation Scenarios

There are three major scenarios employed in the simulations: A, B and C. These scenarios generally differ in the placement of the RIS, UE and BS as shown in Fig. 3. Scenarios B and C also have minor variations; without an explicit statement of the use of a minor variation, it may be assumed that the major scenario is observed. For Scenario A, the BS location is fixed, the relative position of the UE and RIS is fixed, whilst the latitudinal positions of the UE and RIS are simultaneously changed. For Scenario B, the locations of the BS and UE are fixed whilst the latitudinal position of the RIS is changed. A minor variation of Scenario B, is the RIS fixed as shown in Fig. 3 whilst the UE's position is changed longitudinally only. For Scenario C, the BS and RIS positions are fixed; the UE's position is confined to an area of 100λ-by-100λ, where λ is the transmission wavelength. As shown in Fig. 3, the user is not in line-of-sight (LOS) of the BS.

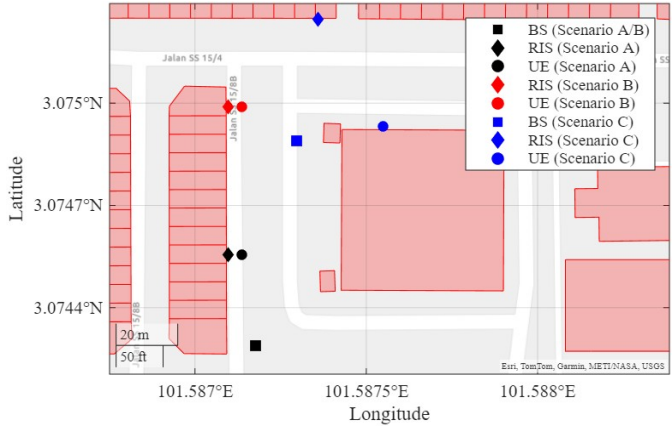

Fig. 3 Placement of BS, RIS and UE for the three major scenarios used in simulation.

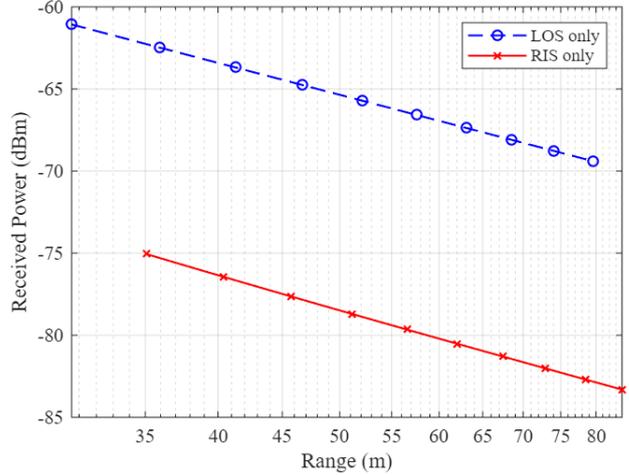

Fig. 4 Free space loss in Scenario A.

## 3 Results and Discussion

The primary motivation of this work is to study the RIS in a multipath propagation environment. Before launching into this area, the simulations in relatively simpler environments where RIS theory have been developed and where wireless propagation behaviour is well-known are investigated first in the free-space propagation and two-ray ground reflection subsections below. Following these, the full multipath propagation investigation is presented with the contribution of reflected rays from buildings and the ground considered. In the last subsection, the non-LOS (NLOS) propagation with reflection and diffraction modes are examined.

### 3.1 Free-Space Propagation

The reflection parameter of the ray tracer was set to zero, hence generating free-space rays only from the BS or RIS. The result of the simulations are shown in Fig. 4 and Fig. 5. The free space loss is observed to increase by 6 dB per octave as expected. The RIS rays travel an additional 4.5m but this does not fully account for the 14 dB additional loss compared to the direct LOS path. This loss is caused by the active element of the RIS which is implicitly modelled by the cascade in the second term of (1) and having the rays of the RIS path segmented into two independent rays, one from the BS to the RIS and the other from the RIS to the UE.

The received power variation as the RIS position is changed in Scenario B is shown in Fig. 5. A minor variation for this scenario is made – the orthogonal distance of the UE from the wall to the east is set identical to the orthogonal distance of the BS from the wall. The result shows that the RIS is ideally placed close to the BS, the second best spot being close to the UE. This result agrees in general with the example given in [8] but differs from it due to the difference in antenna heights.

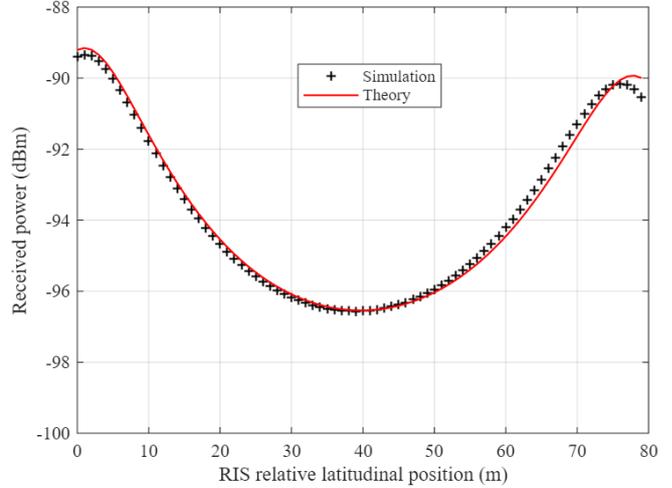

Fig. 5 Received power variation as RIS position is changed.

Applying the same approach given in [8], the received power satisfies

$$P_r \propto \frac{1}{(4\pi d_r)^2 (4\pi d_t)^2} \quad (3)$$

where $d_r$ is the RIS-to-UE distance and $d_t$ is the BS-to-RIS distance. The solid red line in Fig. 5 is obtained using (3) with a proportionality constant of approximately 15 obtained by trial and error.

## 3.2 Two-Ray Ground Reflection

The reflection parameter of the ray tracer was set to one, allowing rays to be generated that can bounce off a wall or the ground at most once. However, in this subsection all rays that bounce off a wall are filtered out and only two rays are kept, the direct LOS ray and the ray that is reflected off the ground. This allows comparison of the results with the theory of two-ray ground reflection that are well-known.

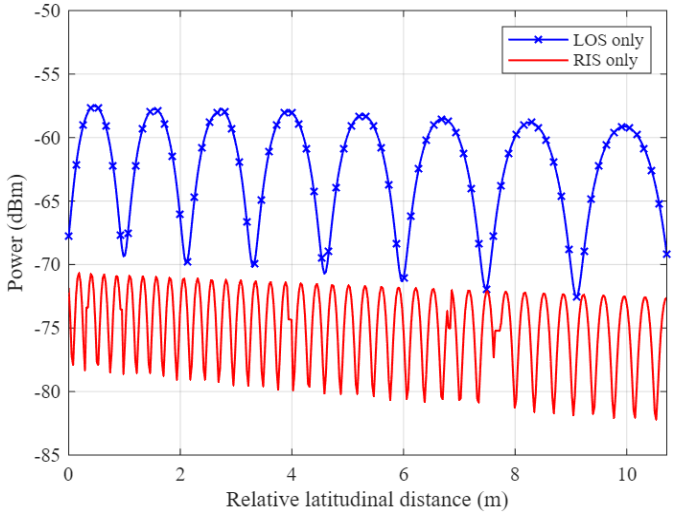

Fig. 6 Variation in power for Scenario A with only the direct ray and ground reflected ray kept for each path segment.

The variation in power for the LOS path, in Scenario A, shown in Fig. 6 is the expected result from the classical theory. The spacing between the nulls increase monotonically and the peaks decrease at the free-space rate. The resultant plot labelled "RIS only", is caused by a double ground reflection and the phase shift due to the RIS. The first ground reflection occurs from the BS to the RIS and the second ground reflection occurs from the RIS to the UE. The incident angles of these reflections are not the same due to the fixed and shorter distance of the UE from the RIS. The anomalous changes of the RIS plot from the picket fence waveform is caused by the RIS changing the phase according to (2) and the varying reflection in the first leg of the RIS channel path. It can also be observed that there are more rapid variations in the RIS path compared to the LOS path. This is caused by the product of the sinusoidal dependence of the total electric field magnitude [22] in each RIS segment.

## 3.3 Multipath Propagation

The results in the following subsections were obtained with the reflection parameter of the ray tracer set to five, i.e. rays may reflect up to five times in the environment.

### 3.3.1 Scenario A

For the simulation results in this section, the latitudes of the UE and RIS were simultaneously incremented in steps of $(2.5 \times 10^{-6})°$ until a total change of $(1 \times 10^{-5})°$ was affected, whilst keeping their longitudes unchanged. This had the effect of increasing the latitudinal distance of the RIS/UE up to 10.6m.

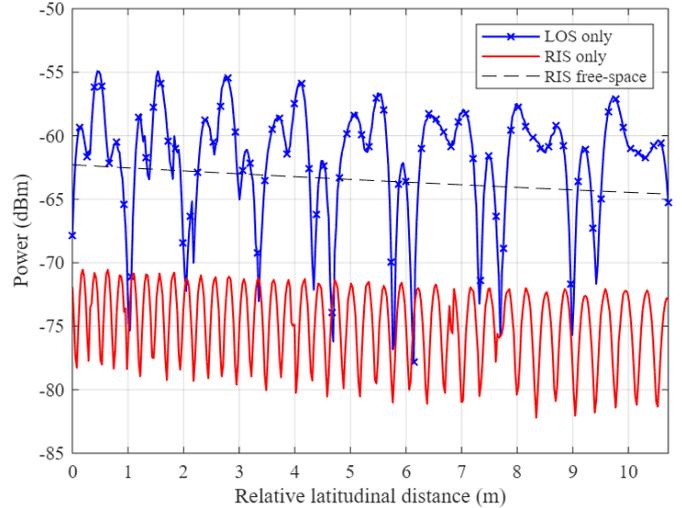

Fig. 7 Effect of multipath fading on RIS channel.

The power received by the UE is shown in Fig. 7. Significantly more variations due to additional reflections from buildings can be seen in the LOS link compared to the same link in Fig. 6 where only the ground reflection affects the total received signal. At certain distances, the signal is boosted by a few dB relative to the peaks of the pickets in Fig. 6 but the fades (nulls) are also deeper.

The received power at these deep fade positions are comparable to the power received due to the RIS links alone. However the effect is more pronounced for multipath propagation in comparison with ground reflection only; it is even possible for the signal from the RIS to be stronger than the LOS path in the deep fades. The power variations in the RIS link are clearly dominated by the ground reflection components with the additional reflections causing minor variations in the peaks of the picket. The dashed line in Fig. 7 is due to free space propagation for the total RIS path but without the RIS showing a 9 dB penalty, relative to the peaks, when the RIS is applied.

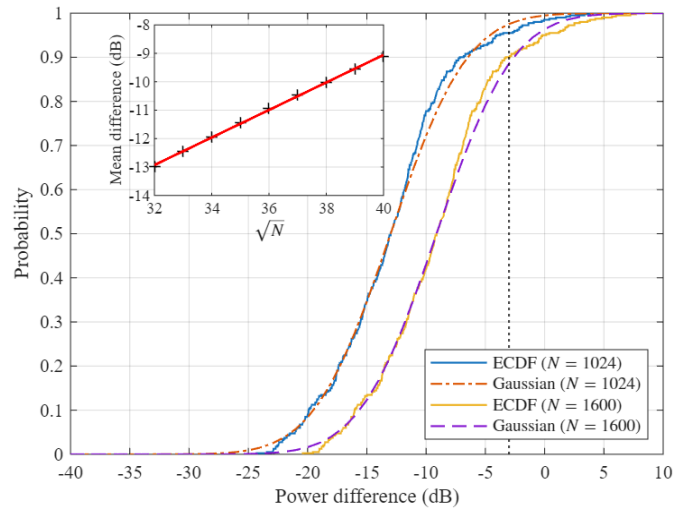

Fig. 8 Empirical cumulative distribution functions of the difference in received power.

Whilst the ray tracing is a deterministic computation, the power variation is difficult to understand analytically because of multiple contributions from numerous reflections. The symbols passed to the QPSK modulator are pseudorandom with mean difference in the results of approximately 0.5 dB across simulation runs. The variations in power were never observed to exceed 1 dB across runs due to the (pseudo-)randomness of the source.

Primarily due to analytical complexity, the power received in the LOS and RIS links are treated as random variables. The empirical cumulative distribution functions (ECDF) of the difference of these random variables are shown in Fig. 8 for two RIS sizes. For each of the empirical curves in Fig. 8, a Gaussian CDF has also been fitted to the data with reasonable fits at the low ends, deviating towards the tails especially at the top ends where the difference is underestimated. At the bottom end, there is a finite limit in the power difference that is governed by the total losses from free-space propagation, reflection losses and RIS-induced loss.

With a RIS size of 1024, there is a 4% probability that the difference in power is 3 dB (dotted in line in Fig. 8) or smaller and a 1% chance that the RIS channel has better reception. This probability rises to 10% and 5% respectively, when the RIS size is increased to 1600. For the first RIS size, there is 50% probability that the difference is 13 dB or smaller. For the larger RIS, the difference reduces to 9 dB (or smaller) at 50% probability. The inset in Fig. 8 shows the variation in the mean difference with respect to the RIS size. The absolute difference increases at a rate of about 0.5 dB for every additional row and column added to the RIS.

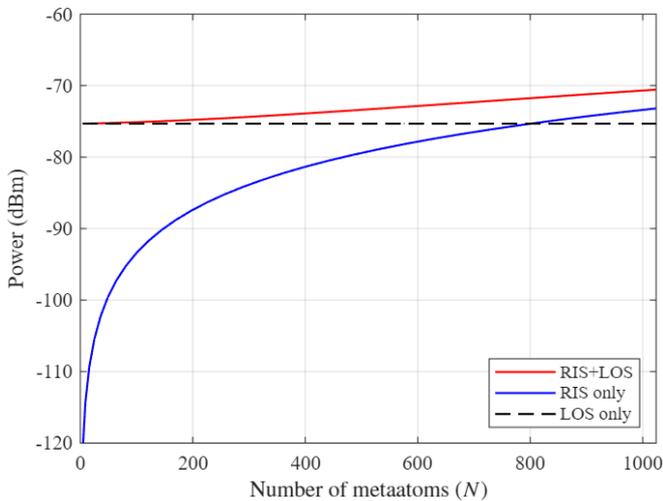

Fig. 9 Channel performance at DF1 with varying RIS size.

A secondary simulation with the RIS and UE latitude fixed at the deep fade near the 1m distance (DF1) as shown in Fig. 7, was conducted with the number of elements in the RIS varied. The result of this second experiment is shown in Fig. 9 which is in excellent agreement with theory [8]. However, the numerical results are strongly dependent on the spatial arrangement of the gNB, RIS and UE, i.e. their locations in the environment. This particular simulation was repeated at similar deep locations where the RIS performed well and the results were similar to that shown for DF1 in Fig. 9.

In Fig. 10 the results of repeating the experiment just described, where in addition to the optimal RIS coefficients obtained from equation (2), the result of using unit coefficients and randomised coefficients with fixed unit magnitude and angles between zero and $2\pi$ picked from a uniform distribution, are shown. The power received at the UE shown in Fig. 10 is the contribution from the RIS channel only and does not include the contribution from the LOS channel.

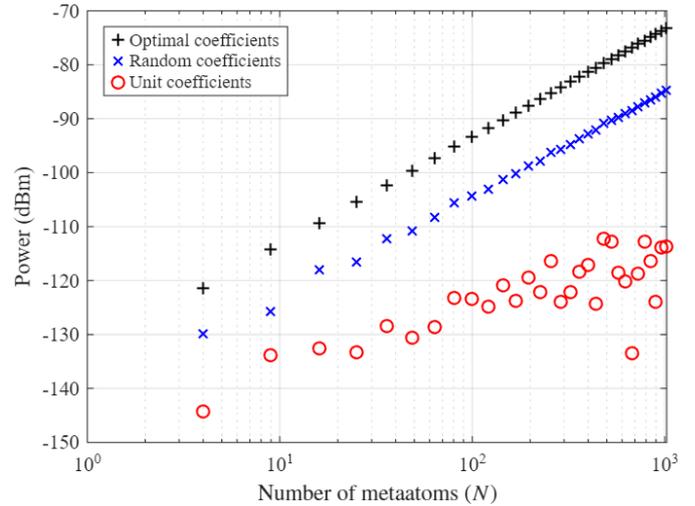

Fig. 10 Effect of the coefficients on the RIS channel.

The result demonstrates the superior performance of the RIS channel with optimal coefficients, with the improvement satisfying an $N^2$ law [8]. A similar result with the one shown here can be found in [24]. Using uniformly distributed RIS coefficient angles, the RIS acts approximately like a diffuse scatterer, and is a reasonable starting point for RIS coefficient optimization algorithms [8], [25]; there is a chance that one of the scattered signals will hit the intended recipient. For the simulations conducted in this section, the optimal performance is about 10 dB away from the performance using randomised coefficients. For the unit coefficients, it was observed that the product of the incident channel angles and the reflected channel angles had a very small probability of connecting with the UE.

### 3.3.2 Scenario A With Diffraction

The simulation of Scenario A is repeated with the diffraction parameter of the ray tracer set to unity and the results are shown in Fig. 11. For the LOS link, the inclusion of diffracted rays improves reception by a few dB and reduces the losses due to deep fades at certain locations. At the 9m position, the deep fade is reduced by about 8dB. The effect of including diffracted rays on the RIS channel is negligible. As a consequence, using the power difference used in Section 3.3.1 for the ECDF as a figure of merit, the performance of the RIS is diminished.

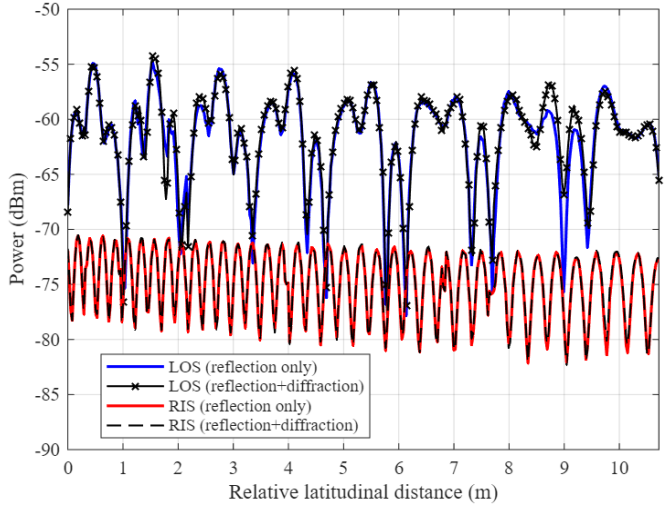

Fig. 11 Effect of multipath fading including diffracted rays.

*3.3.3 Scenario B*

Two experiments were conducted for this scenario. In the first, the RIS is moved along a latitude, away from the UE and towards the gNB. The gNB and UE remain stationary. The result of this experiment is shown in Fig. 12. For the second simulation, the UE instead is moved and the other configuration elements remain stationary. The results are shown in Fig. 13. The simulations in this subsection are not concerned with Doppler effects and the speed of the UE, but are focussed on the relative positions of the RIS, UE and gNB, i.e. the configuration. In each experiment, the results shown are contributed by the RIS channel only.

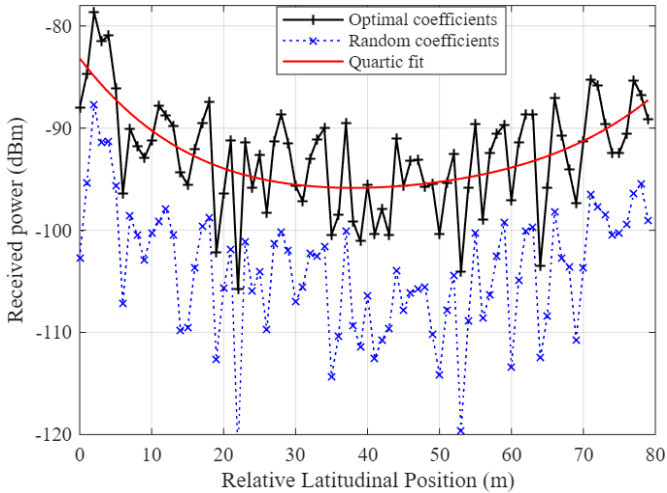

Fig. 12 Received power as RIS moved towards the UE.

The plot shown in Fig. 12 for the RIS with optimal coefficients and multipath propagation is for multiple reflections (up to five) in the environment. If the allowed reflections are restricted to the strongest ground reflections only, i.e. the contributions from all other reflections are discarded, the result for the optimal coefficients is very close to that shown in Fig. 12. As in the previous subsection, ground reflection components dominate the RIS channel performance. However, the picket fence effect is no longer seen because unlike the last section, both legs of the RIS channel path are varying whereas in section 3.3.1 only the first leg, from the BS to the RIS, was varying whilst the second leg, from the RIS to the UE was fixed.

In Fig. 12, the received power contributed by the RIS is strongest when it is either close to the UE or close to the gNB with an advantage for placing the RIS closer to the latter. This result is not unlike that observed for free-space propagation and the same scenario as depicted in Fig. 5. The average variation on top of the multipath fading variations has a U-shape characteristic. According to [2], [8] and (3), the channel gain is essentially determined by a fourth-degree polynomial with the UE-RIS distance as the variable. By linearity, this is also the case here on average as shown by the quartic polynomial fitted to the curve for the optimal coefficients with good agreement.

The model used above is valid when the distance of the UE and BS from the wall is fixed. The channel gain also depends on the distance of the UE from the RIS. In the second experiment for this subsection, the UE is moved away from the RIS and the latter is fixed. When the UE is far from the RIS, the received power follows the free-space path loss in addition to the multipath variations as shown in Fig. 13.

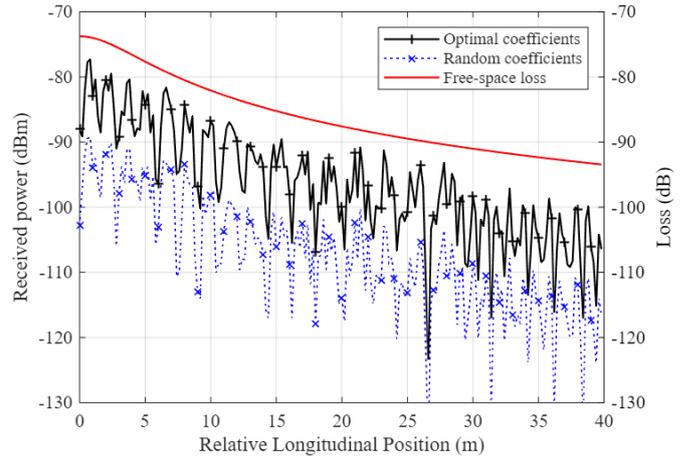

Fig. 13 Received power as UE moved away from RIS.

*3.4 Non-Line-of-Sight Paths*

The simulations presented in the following subsections use the configuration of Scenario C where the UE is not in LOS of the BS; the UE is still in LOS of the RIS. The number of metaatoms $N$ of the RIS was set to 1600. All simulations in this work were done using an AMD 7700 CPU at factory clock speeds. Although, an Nvidia RTX 2080 GPU was available for ray tracing, the computational speed improvement was seen in a handful of observations to speed up by a factor of 1.2x at best in favour of the GPU. With reflections only, simulation times were on the order of a few minutes. When diffracted rays are enabled, the time taken to complete the various simulations increased exponentially. Total system RAM was 32 GB with 27 GB available for simulation and 4-5 GB reserved for the operating system. The orthogonal spacing of the UEs was set to 2.5λ for a total 1681 UE positions in a 41x41 rectangular mesh.

### 3.4.1 Scenario C with Reflections Only

The reflection parameter of the ray tracer was set to five and diffraction disabled. In Fig. 14 and Fig. 15, the rays and power received are shown without and with the aid of the RIS, respectively. In both Fig. 14(a) and Fig. 15(a), the tear-shaped blob is the position of the BS and the view is from an elevation facing eastwards.

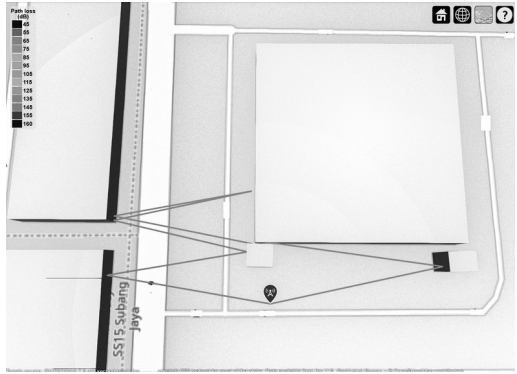

(a) Reflected rays

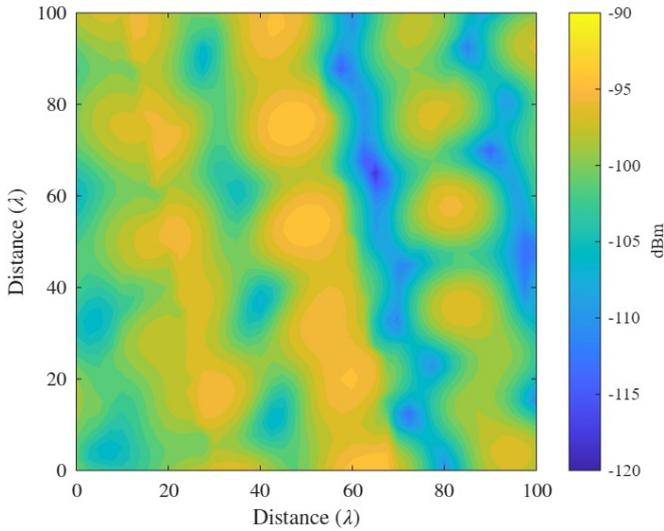

(b) Received power

Fig. 14 Reception without the aid of the RIS.

Fig. 14(a) shows a few of the rays that were calculated and able to reach the non-LOS user without the RIS. The ray travelling north-north-east is reflected thrice off nearby buildings; the ray travelling south-south-east is reflected four times passing through the narrow gap between the small structure and the larger building. The reception interference pattern due to all the rays is shown in Fig. 14(b). There is a general orientation of the pattern due to the angle of arrival of the rays as reflected from the wall to the north of the UE. The signal received is contributed mainly by multiple reflections off buildings.

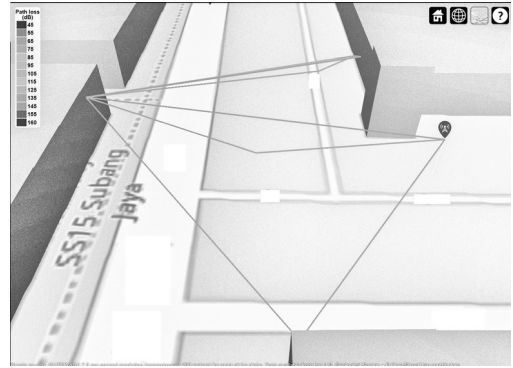

(a) Reflected rays

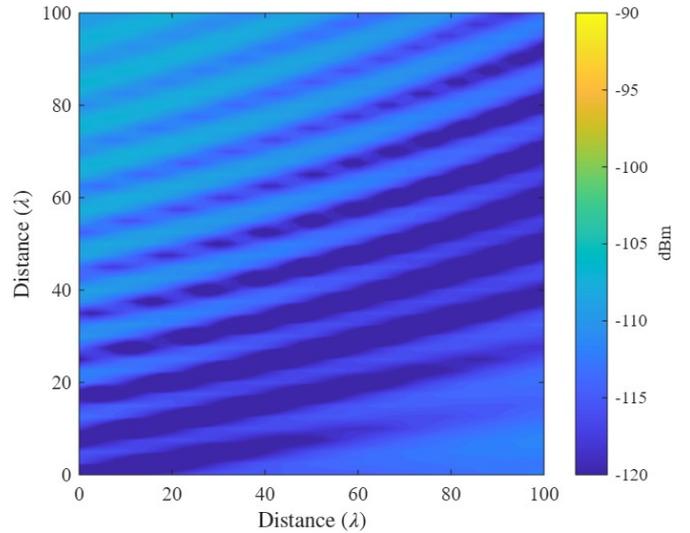

(b) Received power

Fig. 15 Reception with rays emanating from the RIS only.

In Fig. 15(a) a few of the rays emanating from the RIS arrive at the UE via two ground reflections and two wall reflections in addition to the free-space paths. As shown in Fig. 15(b), the signal is about 15 dB weaker on average compared to Fig. 14(b). The interference pattern is due to the ground reflection and the angle of the wavefronts is in good agreement with the angle of arrival of the rays as shown in Fig. 15(a). It should be noted that the ground reflection incident angles are closer to grazing angles compared to the acute angles in Fig. 14(a). Furthermore, the complex permittivities of the building and ground are not identical. Thus, it is not possible to attribute the 15 dB difference due to the RIS alone.

### 3.4.2 Scenario C with Reflection and Diffraction

The diffraction parameter was set to unity and the reflection parameter was kept at five as in the previous subsection. The simulation for reception without the aid of the RIS was repeated and conducted in ten batches with each batch taking about 45 minutes to complete. The result in Fig. 16 shows general improvement over Fig. 14(b) and signficant improvement in the deep faded areas similar to the result in Fig. 11.

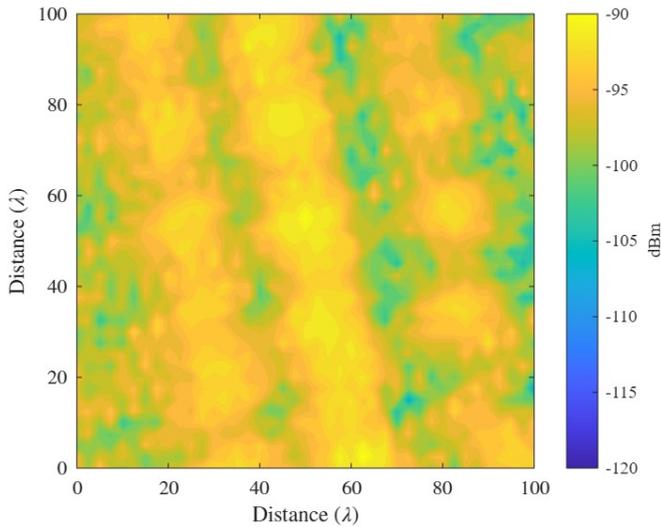

Fig. 16 Reception with reflection and diffraction but no RIS.

## 4 Conclusion

Ray tracing is a viable method for investigating RIS-assisted channels in a multipath environment. The computational requirements are modest when a small number of reflections is enabled for the ray tracing function. The effectiveness of a RIS in multipath fading depends on the configuration of the RIS, gNB and the UE. In the best case, when the LOS link is weak and the RIS channel is relatively stronger, with enough RIS elements, employing a RIS is beneficial especially if the LOS link were to weaken significantly for a brief moment.